\documentclass[twoside,a4paper]{amsart}
\usepackage{graphicx}
\PassOptionsToPackage{pdfauthor={Vladimir V. Kisil},%
    pdftitle={Hypercomplex Representations of the Heisenberg Group and Mechanics},%
    pdfsubject={physics},%
    backref=page,%
  pdfkeywords={symmetries, quantum mechanics, classical mechanics,
    non-standard analysis, dual numbers, double numbers}
}{hyperref}
\IfFileExists{eulervm.sty}{\usepackage{eulervm}
}{}

\usepackage[printedin,extnum]{myamsart}
\ppnum{%LEEDS--MATH--PURE--200?-??
}{arXiv: \href{http://arXiv.org/abs/1005.5057}{arXiv:1005.5057}}
{2010}

\input{cyr.fd}

\providecommand{\rmh}{\mathrm{j}}
\input{mydef}
\providecommand{\myh}{h}
\providecommand{\myhbar}{\hslash}

\providecommand{\anti}{\mathcal{A}}
\providecommand{\ub}[3][]{\left\{\!#1\left[#2,#3\right]\!#1\right\}}

\providecommand{\keywords}[1]{}
\providecommand{\urladdr}[1]{}

\begin{document}
\title[EPAL3.1 Hypercomplex Representations of \(\Space{H}{n}\)
and Mechanics]{Erlangen Programme at Large 3.1\\ 
  Hypercomplex Representations of the Heisenberg Group and Mechanics}

\author[Vladimir V. Kisil]%
{\href{http://www.maths.leeds.ac.uk/~kisilv/}{Vladimir V. Kisil}}
\thanks{On  leave from Odessa University.}

\address{%            
%Institute of Mathematics\\
%Economics and Mechanics\\
%Odessa State University\\
%ul. Petra Velikogo, 2\\
%Odessa-57, 270057, UKRAINE
School of Mathematics\\
University of Leeds\\
Leeds LS2\,9JT\\
UK
}

\email{\href{mailto:kisilv@maths.leeds.ac.uk}{kisilv@maths.leeds.ac.uk}}

\urladdr{\href{http://www.maths.leeds.ac.uk/~kisilv/}%
{http://www.maths.leeds.ac.uk/\~{}kisilv/}}

\date{27th May 2010}
\dedicatory{Dedicated to the memory of V.I.~Arnold}
\begin{abstract}
  In the spirit of geometric quantisation we consider representations
  of the Heisenberg(--Weyl) group induced by hypercomplex characters
  of its centre.  This allows to gather under the same framework,
  called p-mechanics, the three principal cases: quantum mechanics
  (elliptic character), hyperbolic mechanics and classical mechanics
  (parabolic character).  In each case we recover the corresponding
  dynamic equation as well as rules for addition of probabilities.
  Notably, we are able to obtain whole classical mechanics
  without any kind of semiclassical limit \(\myhbar\rightarrow 0\).
\end{abstract}
\AMSMSC{81R05}{81R15, 22E27, 22E70, 30G35, 43A65}
\maketitle
\tableofcontents
\section{Introduction}
\label{sec:introduction}
%  This is a report on the ongoing work on unification of quantum and
%  classical mechanical formalism, known as p-mechanics. 

Complex valued representations of the Heisenberg group (also known as
Weyl or Heisenberg-Weyl group) provide a natural
framework for quantum mechanics~\cites{Howe80b,Folland89}. This is the
most fundamental example of the Kirillov orbit method and geometrical
quantisation technique~\cites{Kirillov99,Kirillov94a}.

Following the pattern we consider representations of the Heisenberg
group which are induced by hypercomplex characters of its centre.
Besides complex numbers (which correspond to the elliptic case) there
are two other types of hypercomplex numbers: dual (parabolic) and
double (hyperbolic)~\citelist{\cite{Yaglom79}*{App.~C} \cite{Kisil09c}}.

To describe dynamics of a physical system we use a universal equation
based on inner derivations of the convolution
algebra~\citelist{\cite{Kisil00a} \cite{Kisil02e}}.  The complex
valued representations produce the standard framework for quantum
mechanics with the Heisenberg dynamical equation~\cite{Vourdas06a}.

The double number valued representations, with the hyperbolic unit
\(\rmh^2=1\), is a natural source of hyperbolic quantum mechanics
developed for a
while~\cites{Hudson04a,Hudson66a,Khrennikov03a,Khrennikov05a,Khrennikov08a}.
The universal dynamical equation employs hyperbolic commutator in this
case. This can be seen as a Moyal bracket based on the hyperbolic sine
function. The hyperbolic observables act as operators on a Krein space
with an indefinite inner product. Such spaces are employed in study of
\(\mathcal{PT}\)-symmetric Hamiltonians and hyperbolic
unit \(\rmh^2=1\) naturally appear in this
setup~\cite{GuentherKuzhel10a}. 

The representations with values in dual numbers provide a convenient
description of the classical mechanics. For this we do not take any sort
of semiclassical limit, rather the nilpotency of the parabolic unit
(\(\rmp^2=0\)) do the task. This removes the vicious necessity to
consider the Planck \emph{constant} tending to zero. %The mathematical model
% uses R-min (R-max) algebras, which are studied within tropical
% mathematics and Maslov dequantisation.
The dynamical equation takes the Hamiltonian form. We also describe
classical non-commutative representations of the Heisenberg group
which acts in the first jet space.

\begin{rem}
  \label{re:quantum-complex}
  It is commonly accepted that the striking difference between quantum
  and classical mechanics is non-commutativity of observables in the
  first case. In particular the Heisenberg commutation relations,
  see~\eqref{eq:heisenberg-comm}, imply the uncertainty principle, the
  Heisenberg equation of motion and other quantum features. However
  our work shows that quantum mechanics is mainly determined by the
  properties of complex numbers. Non-commutative representations of the
  Heisenberg group in dual numbers implies the
  Poisson dynamical equation and local addition of probabilities in
  Section~\ref{sec:class-repr-phase}, which are completely classical. 
\end{rem}

\begin{rem}
  It is worth to note that our technique is different from contraction
  technique in the theory of Lie
  groups~\cites{LevyLeblond65a,GromovKuratov05b}. Indeed a contraction
  of the Heisenberg group \(\Space{H}{n}\) is the commutative
  Euclidean group \(\Space{R}{2n}\) which does not recreate neither
  quantum nor classical mechanics.
\end{rem}

The approach provides not only three different types of dynamics, it
also generates the respective rules for addition of probabilities as well.
For example, the quantum interference is the consequence of the same
complex-valued structure, which directs the Heisenberg equation. The
absence of an interference (a particle behaviour) in the classical
mechanics is again the consequence the nilpotency of the parabolic
unit. Double numbers creates the hyperbolic law of additions of
probabilities which were extensively
investigates~\cites{Khrennikov03a,Khrennikov05a}. There are still
unresolved issues with positivity of the probabilistic interpretation
in the hyperbolic case~\cites{Hudson04a,Hudson66a}. 

The work clarifies foundations of quantum and classical mechanics.  We
recovered from the representation theory the existence of three
non-isomorphic model of mechanics already discussed
in~\cites{Hudson04a,Hudson66a} from translation invariant
formulation. 
It also hinted that hyperbolic counterpart is (at least theoretically)
as natural as classical and quantum mechanics are. The approach
provides a framework for description of aggregate system which have
say both quantum and classical components. This can be used to model
quantum computers with classical terminals~\cite{Kisil09b}.

Remarkably, simultaneously with the work \cite{Hudson66a}
group-invariant axiomatics of geometry lead
R.I.~Pimenov~\cite{Pimenov65a} to description of \(3^n\) Cayley--Klein
constructions. The connection between group-invariant geometry and respective
mechanics were explored in many works of N.A.~Gromov, see for
example~\cites{Gromov90a,Gromov90b,GromovKuratov05b}. Those already
highlighted the r\^ole of three types of hypercomplex units for the
realisation of elliptic, parabolic and hyperbolic geometry and
kinematic.

There is a further connection between representations of the
Heisenberg group and hypercomplex numbers. The symplectomorphism of
phase space are also automorphism of the Heisenberg
group~\cite{Folland89}*{\S~1.2}. Induced representation of the
symplectic group naturally lead to hypercomplex
numbers~\cite{Kisil09c}. Hamiltonians, which produce those
symplectomorphism, are of interest, for example, in quantum
optic~\cite{ATorre10a}. An analysis of those Hamiltonians by means of
creation/annihilation operators recreate hypercomplex coefficients as
well~\cites{Kisil11a,Kisil11c}. 

\begin{rem}
  This work is performed within the ``Erlangen programme at large''
  framework~\cites{Kisil06a,Kisil05a}, thus it would be suitable to
  explain the numbering of various papers. Since the logical order may be
  different from chronological one the following numbering  scheme
  is used:
  \begin{center}
  \begin{tabular}{||c|p{.7\textwidth}||}
    \hline\hline
    Prefix&Branch description\\
    \hline\hline
    ``0'' or no prefix & Mainly geometrical works, within the classical
    field of Erlangen programme by F.~Klein\\
    \hline 
    ``1'' & Papers on analytical functions theories and wavelets\\
    \hline
    ``2'' & Papers on operator theory, functional calculi and
    spectra\\ 
    \hline 
    ``3'' & Papers on mathematical physics\\
    \hline\hline
  \end{tabular}    
  \end{center}
  For example, this is the first paper in the mathematical physics
  area. 
\end{rem}

\section{Heisenberg group and $p$-mechanics}
\label{sec:prel-heis-group}

\subsection{The Heisenberg group and induced representations}
\label{sec:preliminaries}

Let \((s,x,y)\), where \(x\), \(y\in \Space{R}{n}\) and \(s\in\Space{R}{}\), be
an element of the Heisenberg group
\(\Space{H}{n}\)~\cites{Folland89,Howe80b}. The group law on
\(\Space{H}{n}\) is given as follows:
\begin{equation}
  \label{eq:H-n-group-law}
  \textstyle
  (s,x,y)\cdot(s',x',y')=(s+s'+\frac{1}{2}\omega(x,y;x',y'),x+x',y+y'), 
\end{equation} 
where the non-commutativity is due to \(\omega\)---the
\emph{symplectic form} on \(\Space{R}{2n}\)~\cite{Arnold91}*{\S~37}:
\begin{equation}
  \label{eq:symplectic-form}
  \omega(x,y;x',y')=xy'-x'y.
\end{equation}
The Heisenberg group is non-commutative Lie
group with the centre
\begin{displaymath}
  Z=\{(s,0,0)\in \Space{H}{n}, \ s \in \Space{R}{}\}.
\end{displaymath}
The left shifts
\begin{equation}
  \label{eq:left-right-regular}
  \Lambda(g): f(g') \mapsto f(g^{-1}g')  
\end{equation}
act as a representation of \(\Space{H}{n}\) on a certain linear space
of functions. For example, action on \(\FSpace{L}{2}(\Space{H}{},dg)\) with
respect to the Haar measure \(dg=ds\,dx\,dy\) is the \emph{left regular}
representation, which is unitary.

The Lie algebra \(\algebra{h}^n\) of \(\Space{H}{n}\) is spanned by
left-(right-)invariant vector fields
\begin{equation}
\textstyle  S^{l(r)}=\pm{\partial_s}, \quad
  X_j^{l(r)}=\pm\partial_{ x_j}-\frac{1}{2}y_j{\partial_s},  \quad
 Y_j^{l(r)}=\pm\partial_{y_j}+\frac{1}{2}x_j{\partial_s}
  \label{eq:h-lie-algebra}
\end{equation}
on \(\Space{H}{n}\) with the Heisenberg \emph{commutator relations} 
\begin{equation}
  \label{eq:heisenberg-comm}
  [X_i^{l(r)},Y_j^{l(r)}]=\delta_{ij}S^{l(r)} 
\end{equation}
and  all other commutators vanishing. We will omit the supscript \(l\)
for left-invariant field sometimes.

We can construct linear representations by
induction~\cite{Kirillov76}*{\S~13} from a character \(\chi\) of the
centre \(Z\). There are several models for induced representations, here we
prefer the following one, which is presented stripping off all
generalities, cf.~\citelist{\cite{Kirillov76}*{\S~13} 
\cite{MTaylor86}*{Ch.~5}}. Let \(\FSpace[\chi]{F}{2}(\Space{H}{n})\) be
the space of functions on \(\Space{H}{n}\) having the properties:
\begin{equation}
  \label{eq:induced-prop}
  f(gh)=\chi(h)f(g), \qquad \text{ for all } g\in \Space{H}{n},\ h\in Z
\end{equation}
and
\begin{equation}
  \label{eq:L2-condition}
  \int_{\Space{R}{2n}} \modulus{f(0,x,y)}^2dx\,dy<\infty.
\end{equation}
Then \(\FSpace[\chi]{F}{2}(\Space{H}{n})\) is invariant under the left
shifts and those shifts restricted to
\(\FSpace[\chi]{F}{2}(\Space{H}{n})\) make a representation
\(\uir{}{\chi}\) of \(\Space{H}{n}\) induced by \(\chi\).

If the character \(\chi\) is unitary, then the induced representation
is unitary as well. However the representation \(\uir{}{\chi}\) is not
necessarily irreducible. Indeed, left shifts are commuting with the
right action of the group. Thus any subspace of null-solutions of a
linear combination \(aS+\sum_{j=1}^n (b_jX_j+c_jY_j)\) of
left-invariant vector fields is left-invariant and we can restrict
\(\uir{}{\chi}\) to this subspace. The left-invariant differential
operators define analytic condition for functions, cf.~\cite{Vourdas06a}.
\begin{example}
  The function \(f_0(s,x,y)=e^{\rmi \myh s -\myh(x^2 +y^2)/4}\),
  where \(\myh=2\pi\myhbar\), belongs to
  \(\FSpace[\chi]{F}{2}(\Space{H}{n})\) for the character
  \(\chi(s)=e^{\rmi \myh s}\). It is also a null solution for all the
  operators \(X_j-\rmi Y_j\). The closed linear span of functions
  \(f_g=\Lambda(g) f_0\) is invariant under left shifts and provide a
  model for Segal--Bargmann type representation of the Heisenberg
  group, which will be considered below. 
\end{example}

% Note that \(f_\myh(q,p)\) is in
% \(\FSpace{F}{2}(\orbit{\myh})\) if and only if the function
% \(f_\myh(z)e^{-\modulus{z}^2/\myh}\), \(z=p+\rmi q\) is in the
% classical Segal--Bargmann space~\cite{Folland89,Howe80b}, particularly
% is analytical in \(z\). Furthermore the space
% \(\FSpace{F}{2}(\orbit{\myh})\) is spanned by the Gaussian
% \emph{vacuum vector} \(e^{-(q^2+p^2)/\myh}\) and all
% \emph{coherent states}, which are ``shifts'' of the vacuum vector by
% operators~\eqref{eq:stone-inf}.

% Commutative representations~\eqref{eq:stone-one} are oftenly
% forgotten, however their union naturally (see the appearance of
% Poisson bracket in~\eqref{eq:Poisson}) act as the classic
% \emph{phase space}:
% \begin{equation}
%   \label{eq:orbit-0}
%   \orbit{0}=\bigcup_{(q,p)\in\Space{R}{2n}} \orbit{(q,p)}.
% \end{equation}

\begin{rem}
  \label{re:induced-hs}
  An alternative construction of induced representations is as
  follow~\cite{Kirillov76}*{\S~13.2}.  Consider a subgroup \(H\) of a
  group \(G\).  Let a smooth section \(\mathbf{s}:G/H\rightarrow G\)
  be a left inverse of the natural projection
  \(\mathbf{p}:G\rightarrow G/H\). Thus any element \(g\in G\) can be
  uniquely decomposed as \(g=\mathbf{s}(\mathbf{p}(g))*\mathbf{r}(g)\)
  where the map \(\mathbf{r}:G\rightarrow H\) is defined by the previous
  identity. For a character \(\chi\) of \(H\) we can define a
  \emph{lifting} \(\oper{L}_{\chi}: \FSpace{L}{2}(G/H) \rightarrow
  \FSpace[\chi]{L}{2}(G)\) as follows:
  \begin{equation}
    \label{eq:lifting}
    [\oper{L}_{\chi} f](g)=\chi(\mathbf{r}(g))f(\mathbf{p}(g))\qquad
    \text{where } f(x)\in \FSpace{L}{2}(G/H). 
  \end{equation}
  The image space of the lifting \(\oper{L}_{\chi}\) is invariant
  under left shifts.  We also define the \emph{pulling}
  \(\oper{P}:\FSpace[\chi]{L}{2}(G) \rightarrow \FSpace{L}{2}(G/H)\),
  which is a left inverse of the lifting and explicitly cab be given,
  for example, by \([\oper{P}F](x)=F(\mathbf{s}(x))\). Then the
  induced representation on \(\FSpace{L}{2}(G/H)\) is generated by the
  formula \(\uir{}{\chi}(g)=\oper{P}\circ\Lambda (g)\circ\oper{L}\).
\end{rem}

\subsection{Convolutions (observables) on $\Space{H}{n}$ and commutator}
\label{sec:conv-algebra-hg}

Using a left invariant measure \(dg=ds\,dx\,dy\) on \(\Space{H}{n}\) we can
define the convolution of two functions: 
\begin{eqnarray}
  (k_1 * k_2) (g) &=& \int_{\Space{H}{n}} k_1(g_1)\,
  k_2(g_1^{-1}g)\,dg_1  .
  \label{eq:de-convolution}
\end{eqnarray}
This is a non-commutative operation, which is meaningful for functions
from various spaces including \(\FSpace{L}{1}(\Space{H}{n},dg)\), the
Schwartz space \(\FSpace{S}{}\) and many classes of distributions,
which form algebras under convolutions.  Convolutions on
\(\Space{H}{n}\) are used as \emph{observables} in
\(p\)-mechanic~\cites{Kisil96a,Kisil02e}.

A unitary representation \(\uir{}{} \) of \(\Space{H}{n}\) extends
 to \(\FSpace{L}{1}(\Space{H}{n} ,dg)\) by the formula:
\begin{equation}
  \label{eq:rho-extended-to-L1}
  \uir{}{} (k) = \int_{\Space{H}{n}} k(g)\uir{}{}  (g)\,dg .
\end{equation}
This is also an algebra homomorphism of convolutions to linear
operators. 

For a dynamics of observables we need inner \emph{derivations} \(D_k\) of
the convolution algebra \(\FSpace{L}{1}(\Space{H}{n})\), which are
given by the \emph{commutator}:
\begin{eqnarray}
  D_k: f \mapsto [k,f]&=&k*f-f*k   \label{eq:commutator}
   \\  &=&
  \int_{\Space{H}{n}} k(g_1)\left( f(g_1^{-1}g)-f(gg_1^{-1})\right)\,dg_1
, \quad f,k\in\FSpace{L}{1}(\Space{H}{n}).
\nonumber 
\end{eqnarray}

To describe dynamics of a time-dependent observable \(f(t,g)\) we use
the universal equation, cf.~\cites{Kisil94d,Kisil96a}: 
\begin{equation}
  \label{eq:universal}
  S\dot{f}=[H,f],
\end{equation}
where \(S\) is the left-invariant vector
field~\eqref{eq:h-lie-algebra} generated by the centre of
\(\Space{H}{n}\). The presence of operator \(S\) fixes the
dimensionality of both sides of the equation~\eqref{eq:universal} if
the observable \(H\) (Hamiltonian) has the dimensionality of
energy~\cite{Kisil02e}*{Rem~4.1}. If we apply a right inverse
\(\anti\) of \(S\) to both sides of the equation~\eqref{eq:universal}
we obtain the equivalent equation
\begin{equation}
  \label{eq:universal-bracket}
  \dot{f}=\ub{H}{f},
\end{equation}
based on the universal bracket \(\ub{k_1}{k_2}=k_1*\anti k_2-k_2*\anti
k_1\)~\cite{Kisil02e}. 
\begin{example}[Harmonic oscillator]
  \label{ex:p-harmonic}
  Let \(H=\frac{1}{2} (m\omega^2 q^2 + \frac{1}{m}p^2)\) be the
  Hamiltonian of a one-dimensional harmonic oscillator, where
  \(\omega\) is a constant frequency and \(m\) is a constant mass.
  Its \emph{p-mechanisation} will be the second order differential operator
  on \(\Space{H}{n}\)~\cite{BrodlieKisil03a}*{\S~5.1}:
  \begin{displaymath}
    \textstyle
    H=\frac{1}{2} (m\omega^2 X^2
    + \frac{1}{m}Y^2),
  \end{displaymath}
  where we dropped sub-indexes of vector
  fields~\eqref{eq:h-lie-algebra} in one dimensional setting. We can
  express the commutator as a difference between the left and the
  right action of the vector fields:
  \begin{displaymath}
    \textstyle
    [H,f]=\frac{1}{2} (m\omega^2 ((X^{r})^2-(X^{l})^2)
    + \frac{1}{m}((Y^{r})^2-(Y^{l})^2))f.
  \end{displaymath}
   Thus the equation~\eqref{eq:universal} becomes~\cite{BrodlieKisil03a}*{(5.2)}:
   \begin{equation}
     \label{eq:p-harm-osc-dyn}
     \frac{\partial }{\partial s}\dot{f}= \frac{\partial }{\partial s}
    \left(m \omega^2 y\frac{\partial}{\partial x}-\frac{1}{m} x
      \frac{\partial}{\partial y} \right) f. 
   \end{equation}
   Of course, the derivative \(\frac{\partial }{\partial s}\) can be
   dropped from both sides of the equation and the general solution
   is found to be:
   \begin{equation}
     \label{eq:p-harm-sol}
     \textstyle
     f(t;s,x,y)  =  f_0\left(s, x\cos(\omega
    t) +
      m \omega y\sin( \omega t), 
      -\frac{x}{m\omega} \sin(\omega t) + y\cos (\omega t)\right),
   \end{equation}
   where \(f_0(s,x,y)\) is the initial value of an observable on \(\Space{H}{n}\).
\end{example}
\begin{example}[Unharmonic oscillator] 
  \label{ex:p-unharmonic}
  We consider unharmonic
  oscillator with cubic potential, see~\cite{CalzettaVerdaguer06a} and
  references therein:
  \begin{equation}
    \label{eq:unharmonic-hamiltonian}
    H=\frac{m\omega^2}{2} q^2+\frac{\lambda}{6} q^3
  + \frac{1}{2m}p^2.
  \end{equation}
  Due to absence of non-commutative products p-mechanisation is straightforward: 
  \begin{displaymath}
    H=\frac{m\omega^2}{2}  X^2+\frac{\lambda}{6} X^3
    + \frac{1}{m}Y^2.
  \end{displaymath}
  Similarly to the harmonic case the dynamic equation, after
  cancellation of \(\frac{\partial }{\partial s}\) on both sides,
  becomes:
   \begin{equation}
     \label{eq:p-unharm-osc-dyn}
     \dot{f}=     \left(m \omega^2 y\frac{\partial}{\partial x}
       +\frac{\lambda}{6}\left(3y\frac{\partial^2}{\partial x^2} 
         +\frac{1}{4}y^3\frac{\partial^2}{\partial s^2}\right)-\frac{1}{m} x
      \frac{\partial}{\partial y} \right) f. 
   \end{equation}
\end{example}
Unfortunately, it cannot be solved analytically as easy as the
harmonic case.
\subsection{States and Probability}
\label{sec:states-probability}

Let an observable \(\uir{}{}(k)\)~\eqref{eq:rho-extended-to-L1} is
defined by a kernel \(k(g)\) on the Heisenberg group and its
representation \(\uir{}{}\) at a Hilbert space \(\mathcal{H}\). A
state on the convolution algebra is given by a vector
\(v\in\mathcal{H}\). A simple calculation:
\begin{eqnarray*}
  \scalar[\mathcal{H}]{\uir{}{}(k)v}{v}&=& \scalar[\mathcal{H}]{\int_{\Space{H}{n}} k(g)
    \uir{}{}(g)v\,dg}{v}\\
  &=& \int_{\Space{H}{n}} k(g) \scalar[\mathcal{H}]{\uir{}{}(g)v}{v}dg\\
  &=& \int_{\Space{H}{n}} k(g) \overline{\scalar[\mathcal{H}]{v}{\uir{}{}(g)v}}\,dg
\end{eqnarray*}
can be restated as:
\begin{displaymath}
  \scalar[\mathcal{H}]{\uir{}{}(k)v}{v}=\scalar[]{k}{l}, \qquad \text{where} \quad
  l(g)=\scalar[\mathcal{H}]{v}{\uir{}{}(g)v}.
\end{displaymath}
Here the left-hand side contains the inner product on
\(\mathcal{H}\), while the right-hand side uses a skew-linear pairing
between functions on \(\Space{H}{n}\) based on the Haar measure
integration. In other words we obtain,
cf.~\cite{BrodlieKisil03a}*{Thm.~3.11}:
\begin{prop}
  \label{pr:state-functional}
  A state defined by a vector \(v\in\mathcal{H}\) coincides with the
  linear functional given by the wavelet transform
  \begin{equation}
    \label{eq:kernel-state}
    l(g)=\scalar[\mathcal{H}]{v}{\uir{}{}(g)v}
  \end{equation}
  of \(v\) used as the mother wavelet as well.
\end{prop}
The addition of vectors in \(\mathcal{H}\) implies the following
operation on states:
\begin{eqnarray}
  \scalar[\mathcal{H}]{v_1+v_2}{\uir{}{}(g)(v_1+v_2)}&=&
    \scalar[\mathcal{H}]{v_1}{\uir{}{}(g)v_1}
    +\scalar[\mathcal{H}]{v_2}{\uir{}{}(g)v_2}\nonumber \\
    &&{}+%2\Re 
    \scalar[\mathcal{H}]{v_1}{\uir{}{}(g)v_2}
   + \overline{\scalar[\mathcal{H}]{v_1}{\uir{}{}(g^{-1})v_2}}
   \label{eq:kernel-add}
\end{eqnarray}
The last expression can be conveniently rewritten for kernels of the
functional as 
\begin{equation}
  \label{eq:addition-functional}
  l_{12}=l_1+l_2+2 A\sqrt{l_1l_2}
\end{equation}
for some real number \(A\). This formula is behind the contextual law
of addition of conditional probabilities~\cite{Khrennikov01a} and will
be illustrated below. Its physical interpretation is an interference, say, from
two slits. The mechanism of such interference can be both
causal  and local, see~\citelist{\cite{Kisil01c} \cite{KhrenVol01}}.

\section{Elliptic characters and Quantum Dynamics}
\label{sec:ellipt-char-moyal}
In this section we consider the representation \(\uir{}{\myh}\) of
\(\Space{H}{n}\) induced by the elliptic character
\(\chi_\myh(s)=e^{\rmi\myh s}\) in complex numbers parametrised by
\(\myh\in\Space{R}{}\). We also use the convenient agreement \(\myh=2\pi\myhbar\).

\subsection{Segal--Bargmann and Schr\"odinger Representations}
\label{sec:schr-segal-bargm}
The realisation of \(\uir{}{\myh}\) by the left
shifts~\eqref{eq:left-right-regular} on
\(\FSpace[\myh]{L}{2}(\Space{H}{n})\) is rarely used in quantum
mechanics. Instead two unitary equivalent forms are more common: the
Schr\"odinger and Segal--Bargmann representations.

The Segal-Bargmann representation can be obtained from the orbit
method of Kirillov~\cite{Kirillov94a}. It allows spatially separate
irreducible components of the left regular representation, each of
them is located on the orbit of the co-adjoint representation,
see~\citelist{\cite{Kisil02e}*{\S~2.1} \cite{Kirillov94a}} for
details, we only present a brief summary here.

We identify \(\Space{H}{n}\) and its Lie algebra \(\algebra{h}_n\)
through the exponential map~\cite{Kirillov76}*{\S~6.4}. The dual
\(\algebra{h}_n^*\) of \(\algebra{h}_n\) is presented by the Euclidean
space \(\Space{R}{2n+1}\) with coordinates \((\myhbar,q,p)\).  The
pairing \(\algebra{h}_n^*\) and \(\algebra{h}_n\) given by
\begin{displaymath}
  \scalar{(s,x,y)}{(\myhbar,q,p)}=\myhbar s + q \cdot x+p\cdot y.
\end{displaymath}
This pairing defines the Fourier
transform \(\hat{\ }: \FSpace{L}{2}(\Space{H}{n})\rightarrow
\FSpace{L}{2}(\algebra{h}_n^*)\) given by~\cite{Kirillov99}*{\S~2.3}:
\begin{equation}
  \label{eq:fourier-transform}
  \hat{\phi}(F)=\int_{\algebra{h}^n} \phi(\exp X) 
  e^{-2\pi\rmi  \scalar{X}{F}}\,dX \qquad \textrm{ where }
  X\in\algebra{h}^n,\ F\in\algebra{h}_n^*. 
\end{equation}
For a fixed \(\myhbar\) the left regular
representation~\eqref{eq:left-right-regular} is mapped by the Fourier
transform to the Segal--Bargmann type representation
\citelist{\cite{Kisil02e}*{(2.9)} \cite{deGosson08a}*{(1)}}:
\begin{equation}
  \label{eq:stone-inf}
  \textstyle
  \uir{}{\myhbar}(s,x,y): f (q,p) \mapsto 
  e^{-2\pi\rmi(\myhbar s+qx+py)}
  f \left(q-\frac{\myhbar}{2} y, p+\frac{\myhbar}{2} x\right).
\end{equation}

The collection of points \((\myhbar,q,p)\in\algebra{h}_n^*\) for a
fixed \(\myhbar\) is naturally identified with the phase space
of the system.
\begin{rem}
  It is possible to identify the case of \(\myhbar=0\) with classical
  mechanics~\cite{Kisil02e}. Indeed, a substitution of the zero value of \(\myhbar\)
  into~\eqref{eq:stone-inf} produces the commutative representation:
  \begin{equation}
    \label{eq:commut-repres}
      \uir{}{0}(s,x,y): f (q,p) \mapsto 
      e^{-2\pi\rmi(qx+py)}
      f \left(q, p\right).
  \end{equation}
  It can be decomposed into the direct integral of one-dimensional
  representations parametrised by the points \((q,p)\) of the phase
  space. The classical mechanics, including the Hamilton equation, can
  be recovered from those representations~\cite{Kisil02e}. However 
  the condition \(\myhbar=0\) (as well as \(\myhbar\rightarrow 0\)) is
  not completely physical. Commutativity (and subsequent relative
  triviality) of those representation is the main reason why they are
  oftenly neglected. The commutativity can be outweighed by special
  arrangements, e.g. an antiderivative~\cite{Kisil02e}*{(4.1)}, but the
  procedure is not straightforward, see discussion in~\citelist{\cite{Kisil05c}
    \cite{AgostiniCapraraCiccotti07a}
    \cite{Kisil09a}}. A direct approach using dual
  numbers will  be discussed below, cf. Rem.~\ref{re:hamilton-from-nc}.
\end{rem}

To recover the Schr\"odinger representation we use
Rem.~\ref{re:induced-hs}, see~\cite{Kisil98a}*{Ex.~4.1} for details.
The subgroup \(H=\{(s,0,y)\such s\in\Space{R}{},
y\in\Space{R}{n}\}\subset\Space{H}{n}\) defines the homogeneous space
\(X=G/H\), which coincides with \(\Space{R}{n}\) as a manifold. The
natural projection \(\mathbf{p}:G\rightarrow X\) is \(\mathbf{p}(s,x,y)=x\) and its left
inverse \(\mathbf{s}:X\rightarrow G\) can be as simple as \(\mathbf{s}(x)=(0,x,0)\).
For the map \(\mathbf{r}:G\rightarrow H\), \(\mathbf{r}(s,x,y)=(s-xy/2,0,y)\) we have
the decomposition
\begin{displaymath}
  (s,x,y)=\mathbf{s}(p(s,x,y))*\mathbf{r}(s,x,y)=(0,x,0)*(s-\textstyle\frac{1}{2}xy,0,y).
\end{displaymath}
For a character
\(\chi_{\myh}(s,0,y)=e^{\rmi\myh s}\) of \(H\) the lifting
\(\oper{L}_\chi: \FSpace{L}{2}(G/H) \rightarrow
\FSpace[\chi]{L}{2}(G)\) is as follows:
\begin{displaymath}
  [\oper{L}_\chi f](s,x,y)=\chi_{\myh}(\mathbf{r}(s,x,y))\, 
  f(\mathbf{p}(s,x,y))=e^{\rmi\myh (s-xy/2)}f(x).  
\end{displaymath}
Thus the representation \(\uir{}{\chi}(g)=\oper{P}\circ\Lambda
(g)\circ\oper{L}\) becomes:
\begin{equation}
  \label{eq:schroedinger-rep}
  [\uir{}{\chi}(s',x',y') f](x)=e^{-2\pi\rmi\myhbar (s'+xy'-x'y'/2)}\,f(x-x').  
\end{equation}
After the Fourier transform \(x\mapsto q\) we get the Schr\"odinger
representation on the configuration space:
\begin{equation}
  \label{eq:schroedinger-rep-conf}
  [\uir{}{\chi}(s',x',y') \hat{f}\,](q)=e^{-2\pi\rmi\myhbar (s'+x'y'/2)
    -2\pi\rmi x' q}\,\hat{f}(q+\myhbar y').  
\end{equation}
Note that this again turns into a commutative representation
(multiplication by an unimodular function) if \(\myhbar=0\). To get
the full set of commutative representations in this way we need to use the
character \(\chi_{(\myh,p)}(s,0,y)=e^{2\pi\rmi(\myhbar+ py)}\) in the
above consideration. 

\subsection{Commutator and the Heisenberg Equation}
\label{sec:comm-heis-equat}

The property~\eqref{eq:induced-prop} of
\(\FSpace[\chi]{F}{2}(\Space{H}{n})\) implies that the restrictions of
two operators \(\uir{}{\chi} (k_1)\) and \(\uir{}{\chi} (k_2)\) to
this space are equal if
\begin{displaymath}
  \int_{\Space{R}{}} k_1(s,x,y)\,\chi(s)\, ds = \int_{\Space{R}{}} k_2(s,x,y)\,\chi(s)\,ds.
\end{displaymath}
In other words, for a character \(\chi(s)=e^{2\pi\rmi \myhbar s}\) the
operator \(\uir{}{\chi} (k)\) depends only on
\begin{displaymath}
  \hat{k}_s(\myhbar,x,y)=\int_{\Space{R}{}} k(s,x,y)\,e^{-2\pi\rmi \myhbar s}\,ds,
\end{displaymath}
which is the partial Fourier transform \(s\mapsto \myhbar\) of
\(k(s,x,y)\). The restriction to \(\FSpace[\chi]{F}{2}(\Space{H}{n})\)
of the composition formula for convolutions is~\cite{Kisil02e}*{(3.5)}:
\begin{equation}
  \label{eq:composition-ell}
  (k'*k)\hat{_s}%(-\myhbar,x,y) 
  =
  \int_{\Space{R}{2n}} e^{ {\rmi \myh}{}(xy'-yx')/2}\, \hat{k}'_s(\myhbar ,x',y')\,
 \hat{k}_s(\myhbar ,x-x',y-y')\,dx'dy'. 
\end{equation}
Under the Schr\"odinger representation~\eqref{eq:schroedinger-rep-conf} the
convolution~\eqref{eq:composition-ell} defines a rule for composition
of two pseudo-differential operators (PDO) in the Weyl
calculus~\citelist{\cite{Howe80b} \cite{Folland89}*{\S~2.3}}.

Consequently the representation~\eqref{eq:rho-extended-to-L1} of
commutator~\eqref{eq:commutator} depends only on its partial Fourier
transform~\cite{Kisil02e}*{(3.6)}: 
\begin{eqnarray}
  [k',k]\hat{_s}
  &=&   2 \rmi  \int_{\Space{R}{2n}}\!\! \sin(\textstyle\frac{\myh}{2}
  (xy'-yx'))\,\label{eq:repres-commutator}\\
   && \qquad\times 
  \hat{k}'_s(\myhbar, x', y')\,
  \hat{k}_s(\myhbar, x-x', y-y')\,dx'dy'. \nonumber 
\end{eqnarray}
Under the Fourier transform~\eqref{eq:fourier-transform} this
commutator is exactly the Moyal bracket~\cite{Zachos02a} for of
\(\hat{k}'\) and \(\hat{k}\).

For observables in the space
\(\FSpace[\chi]{F}{2}(\Space{H}{n})\) the action of \(S\) is reduced to
multiplication% by the derivative of the character
, e.g. for \(\chi(s)=e^{\rmi \myh s}\) the action of \(S\) is
multiplication by \(\rmi \myh\). Thus the
equation~\eqref{eq:universal} reduced to the space
\(\FSpace[\chi]{F}{2}(\Space{H}{n})\) becomes the Heisenberg type
equation~\cite{Kisil02e}*{(4.4)}:
\begin{equation}
  \label{eq:heisenberg-eq}
  \dot{f}=\frac{1}{\rmi\myh}  [H,f]\hat{_s},
\end{equation}
based on the above bracket~\eqref{eq:repres-commutator}.  The
Schr\"odinger representation~\eqref{eq:schroedinger-rep-conf} transforms
this equation to the original Heisenberg equation.

\begin{example}
  \begin{enumerate}
  \item 
    \label{it:q-harmonic}
    Under the Fourier transform \((x,y)\mapsto(q,p)\) the
    p-dynamic equation~\eqref{eq:p-harm-osc-dyn} of the harmonic
    oscillator becomes: 
    \begin{displaymath}
      \dot{f}=     \left(m \omega^2 q\frac{\partial}{\partial p}-\frac{1}{m} p
      \frac{\partial}{\partial q} \right) f. 
    \end{displaymath}
    The same transform creates its solution out
    of~\eqref{eq:p-harm-sol}.
  \item 
    \label{it:q-unharmonic}
    Since \(\frac{\partial}{\partial s}\) acts on
    \(\FSpace[\chi]{F}{2}(\Space{H}{n})\) as multiplication by \(\rmi
    \myhbar\), the quantum representation of unharmonic dynamics
    equation~\eqref{eq:p-unharm-osc-dyn} is:
    \begin{equation}
      \label{eq:q-unhar-dyn}
      \dot{f}=     \left(m \omega^2 q\frac{\partial}{\partial
          p}+\frac{\lambda}{6}\left(3q^2\frac{\partial}{\partial p}
          -\frac{\myhbar^2}{4}\frac{\partial^3}{\partial p^3}\right)-\frac{1}{m}
        p \frac{\partial}{\partial q} \right) f. 
    \end{equation}
    This is exactly the equation for the Wigner function obtained
    in~\cite{CalzettaVerdaguer06a}*{(30)}. 
  \end{enumerate}
\end{example}

\subsection{Quantum Probabilities}
\label{sec:quantum-probabilities}
For the elliptic character \(\chi_\myh(s)=e^{\rmi\myh s }\) we can use
the Cauchy--Schwartz inequality to demonstrate that the real number
\(A\) in the identity~\eqref{eq:addition-functional} is between \(-1\)
and \(1\). Thus we can put \(A=\cos \alpha\) for some angle (phase)
\(\alpha\) to get the formula for counting quantum
probabilities, cf.~\cite{Khrennikov03a}*{(2)}:
\begin{equation}
  \label{eq:addition-functional-ell}
  l_{12}=l_1+l_2+2 \cos\alpha \,\sqrt{l_1l_2}
\end{equation}

\begin{rem}
  \label{re:sine-cosine}
  It is interesting to note that the both trigonometric functions are
  employed in quantum mechanics: sine is in the heart of the Moyal
  bracket~\eqref{eq:repres-commutator} and cosine is responsible for
  the addition of probabilities~\eqref{eq:addition-functional-ell}. In
  the essence the commutator and probabilities took respectively the
  odd and even parts of the elliptic character \(e^{\rmi\myh s}\).
\end{rem}

\begin{example}
%% Details of calculations
% We start from the fundamental identity:
% \begin{displaymath}
%   \int_{\Space{R}{}} e^{-\pi x^2}dx=1
% \end{displaymath}
% and through a change of variables obtain for an \(a>0\):
% \begin{displaymath}
%   \int_{\Space{R}{}} e^{-a x^2+bx+c}dx=\sqrt{\frac{\pi}{a}} \exp\left(\frac{b^2}{4a}+c\right).
% \end{displaymath}
Take a  vector \(v_{(a,b)}\in\FSpace[\myh]{L}{2}(\Space{H}{n})\) defined by a
Gaussian with mean value \((a,b)\) in the phase space for a harmonic oscillator of the
mass \(m\) and the frequency \(\omega\):
\begin{equation}
  \label{eq:gauss-state}
  v_{(a,b)}(q,p)=\exp\left(-\frac{2\pi\omega m}{\myhbar}(q-a)^2-\frac{2\pi}{\myhbar
      \omega m}(p-b)^2\right).
\end{equation}
A direct calculation shows:
\begin{eqnarray*}
  \lefteqn{\scalar{v_{(a,b)}}{\uir{}{\myhbar}(s,x,y)v_{(a',b')}}=\frac{4}{\myhbar}
  \exp\left(
    \pi \rmi \left(2s\myhbar+x (a+a')+y (b+b')\right)\frac{}{}\right.}\\
  &&\left.{} -\frac{\pi}{2 \myhbar\omega m }((\myhbar x+b-b')^2
    +(b-b')^2)
-\frac{\pi\omega m}{2\myhbar} ((\myhbar y+a'-a)^2
  + (a'-a)^2)
  \right)\\
  &=&\frac{4}{\myhbar}
  \exp\left(
    \pi \rmi \left(2s\myhbar+x (a+a')+y (b+b')\right)\frac{}{}\right.\\
  &&\left.{}  -\frac{\pi}{\myhbar\omega m }((b-b'+{\textstyle\frac{\myhbar x}{2}})^2
    +({\textstyle\frac{\myhbar x}{2}})^2)
    -\frac{\pi\omega m}{\myhbar} ((a-a'-{\textstyle\frac{\myhbar y}{2}})^2
    + ({\textstyle\frac{\myhbar y}{2}})^2) 
  \right)
\end{eqnarray*}
Thus the kernel
\(l_{(a,b)}=\scalar{v_{(a,b)}}{\uir{}{\myhbar}(s,x,y)v_{(a,b)}}\)~\eqref{eq:kernel-state}
for a state \(v_{(a,b)}\) is:
\begin{eqnarray}
  l_{(a,b)}&=&\frac{4}{\myhbar}
  \exp\left(
    2\pi \rmi (s\myhbar+xa+yb)\frac{}{}
    -\frac{\pi\myhbar}{2 \omega m }x^2
    -\frac{\pi\omega m \myhbar}{2\myhbar} y^2
  \right)
  \label{eq:single-slit}
\end{eqnarray}
An observable registering a particle at a point \(q=c\) of the
configuration space is \(\delta(q-c)\). On the Heisenberg group this
observable is given by the kernel:
\begin{equation}
  \label{eq:coordinate}
  X_c(s,x,y)=e^{2\pi\rmi (s\myhbar+x  c)}\delta(y).
\end{equation}
The measurement of \(X_c\) on the
state~\eqref{eq:gauss-state} (through the
kernel~\eqref{eq:single-slit}) predictably is: 
\begin{displaymath}
  \scalar{X_c}{l_{(a,b)}}=\sqrt{\frac{2\omega
  m}{\myhbar}}\exp\left(-\frac{2\pi\omega m}{\myhbar}(c-a)^2\right).
\end{displaymath}
\end{example}
\begin{example}
  Now take two states \(v_{(0,b)}\) and \(v_{(0,-b)}\), where for the
  simplicity we assume the mean values of coordinates vanish in the
  both cases.  Then the corresponding kernel~\eqref{eq:kernel-add} has
  the interference terms:
%\begin{eqnarray*}
%  l=l_{(a,0)}+l_{(-a,0)}
%  +  \scalar[\mathcal{H}]{v_1}{\uir{}{}(g)v_2}
%  + \overline{\scalar[\mathcal{H}]{v_1}{\uir{}{}(g^{-1})v_2}}
%\end{eqnarray*}
\begin{eqnarray*}
  l_i&=&  \scalar{v_{(0,b)}}{\uir{}{\myhbar}(s,x,y)v_{(0,-b)}}\\
  &=&\frac{4}{\myhbar}
  \exp\left(2\pi \rmi s\myhbar
    -\frac{\pi}{2 \myhbar\omega m }((\myhbar x+2b)^2
    +4b^2)
    -\frac{\pi\myhbar\omega m}{2} y^2
  \right).
\end{eqnarray*}
The measurement of \(X_c\)~\eqref{eq:coordinate} on this term contains
the oscillating part:
\begin{displaymath}
  \scalar{X_c}{l_i}=\sqrt{\frac{2\omega m}{\myhbar}} \exp\left(-\frac{2\pi\omega
    m }{\myhbar} c^2
  -\frac{2\pi}{\omega
    m \myhbar}b^2+\frac{4\pi\rmi}{\myhbar} cb\right)
\end{displaymath}
Therefore on the kernel \(l\) corresponding to
the state \(v_{(0,b)}+v_{(0,-b)}\) the measurement is 
\begin{eqnarray*}
  \scalar{X_c}{l}%&=&2\sqrt{\frac{2\omega
%      m}{\myhbar}}\exp\left(-\frac{2\pi\omega m}{\myhbar}c^2\right)\\
%  &&{}+2\sqrt{\frac{2\omega m}{\myhbar}} \exp\left(-\frac{2\pi\omega
%      m }{\myhbar} c^2
%    -\frac{2\pi}{\omega
%      m \myhbar}b^2\right)\cos(\frac{4\pi}{\myhbar} cb)\\
  &=&2\sqrt{\frac{2\omega
      m}{\myhbar}}\exp\left(-\frac{2\pi\omega m}{\myhbar}c^2\right)
  \left(1+\exp\left(    -\frac{2\pi}{\omega
      m \myhbar}b^2\right)\cos\left(\frac{4\pi}{\myhbar} cb\right)\right).
\end{eqnarray*}
\begin{figure}[htbp]
  \centering
  (a)\includegraphics[scale=.75]{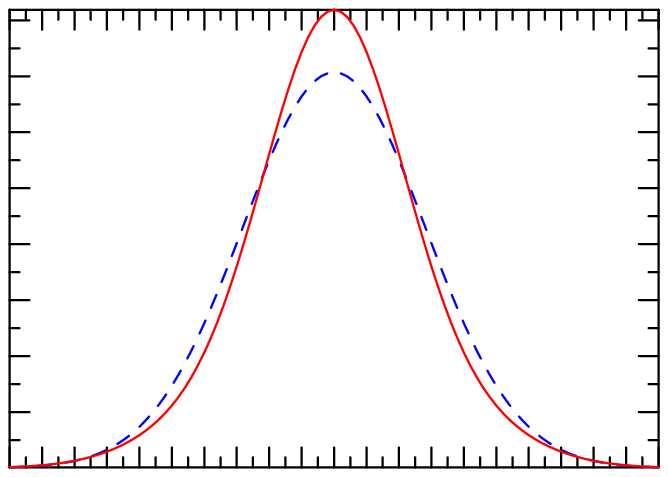}\hfill
  (b)\includegraphics[scale=.75]{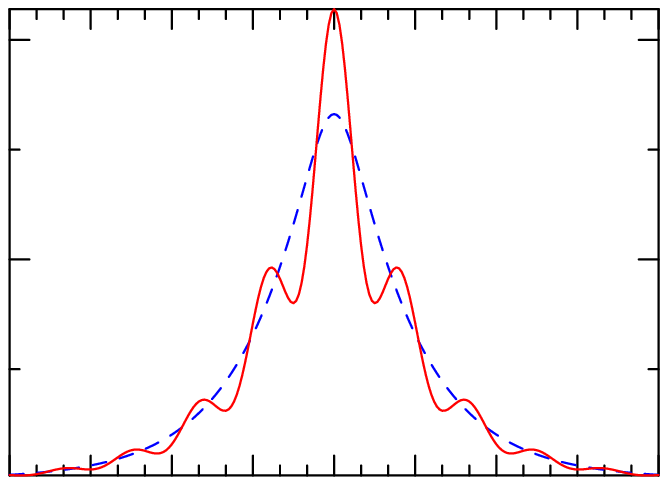}
  \caption{Quantum probabilities: the blue (dashed) graph shows the
    addition of probabilities without interaction, the red (solid)
    graph present the quantum interference. Left picture shows the
    Gaussian state~\eqref{eq:gauss-state}, the right---the rational
    state~\eqref{eq:poly-state}}
  \label{fig:quant-prob}
\end{figure}
The presence of the cosine term in the last expression can generate an
interference picture. In practise it does not happen for the minimal
uncertainty state~\eqref{eq:gauss-state} which we are using here: it
rapidly vanishes outside of the neighbourhood of zero, where
oscillations of the cosine occurs, see Fig.~\ref{fig:quant-prob}(a).
\end{example}

\begin{example}
To see a traditional interference pattern one can use a state which is far
from the minimal uncertainty. For example, we can consider the state:
\begin{equation}
  \label{eq:poly-state}
  u_{(a,b)}(q,p)=\frac{\myhbar^2}{((q-a)^2+\myhbar/\omega
    m)((p-b)^2+\myhbar\omega m)}.
\end{equation}
To evaluate the observable \(X_c\)~\eqref{eq:coordinate} on the
state \(l(g)=\scalar{u_1}{\uir{}{h}(g)u_2}\)~\eqref{eq:kernel-state}
we use the following formula: 
\begin{displaymath}
  \scalar{X_c}{l}=\frac{2}{\myhbar}\int_{\Space{R}{n}} \hat{u}_1(q,
  2(q-c)/\myhbar)\, 
  \overline{\hat{u}_2(q, 2(q-c)/\myhbar)}\,dq,
\end{displaymath}
where \(\hat{u}_{i}(q,x)\) denotes  the partial Fourier transform
\(p\mapsto x\) of \(u_{i}(q,p)\). The formula is obtained by 
swapping order of integrations.  The numerical evaluation of the state
obtained by the addition \(u_{(0,b)}+u_{(0,-b)}\) is plotted on
Fig.~\ref{fig:quant-prob}(b), the red curve shows the canonical
interference pattern. 
\end{example}

\section{Hypercomplex Repersentations of the Heisenberg Group}
\label{sec:hyperc-repers-heis}

The group of symmetries of classical mechanics---the group preserving
the symplectic form~\eqref{eq:symplectic-form}---generates
automorphisms of the Heisenberg group in a natural
way~\cite{Folland89}*{\S~1.2}. Those common symmetries of quantum and
classical mechanics are behind many important connections, e.g.
between classical ``symplectic camel'' and the Heisenberg uncertainty
relations~\cite{deGossonLuef09a}. 

The symplectic group of \(\Space{R}{2}\) is isomorphic to the
celebrated group \(\SL\)~\cite{Lang85}. Both groups \(\Space{H}{n}\)
and \(\SL\) contributes to the symmetries of the paraxial wave
equation~\cite{ATorre10a}. There are many other physical
links between the Heisenberg group and \(\SL\), e.g. metaplectic
representation~\cite{Folland89}*{Ch.~4}. 

It was demonstrated in~\cite{Kisil07a} that dual and double numbers
appears very naturally within the induced representations of the group
\(\SL\). Special relativity~\cite{Ulrych05a} and global space-time
model~\cites{HerranzSantander02b,Kisil06b} also link the
representation theory to hypercomplex numbers.  Physical significance
of hypercomplex numbers and representation theory of Clifford algebras
was recently highlighted as
well~\citelist{\cite{BocCatoniCannataNichZamp07} \cite{Ulrych08a}
  \cite{Plaksa09a} \cite{Ulrych10a}}. There is an explicit similarity
between the commutators in the Heisenberg-Weyl Lie algebra and
anticommutators defining Clifford algebra~\citelist{\cite{Kisil93c}
  \cite{Kisil01d}}, which can be unified as a
superspace~\citelist{\cite{BieEelbodeSommen09a} \cite{Berezin86}}.
Thus it would be an omission to restrict linear representations of
\(\Space{H}{n}\) to complex numbers only.

\subsection{Hyperbolic Representations and Addition of Probabilities}
\label{sec:hyperb-repr-addt}

Now we turn to double numbers also known as hyperbolic, split-complex,
etc. numbers~\citelist{\cite{Yaglom79}*{App.~C} \cite{Ulrych05a}
  \cite{KhrennikovSegre07a}}. They form a two dimensional algebra \(\Space{O}{}\)
spanned by \(1\) and \(\rmh\) with the property \(\rmh^2=1\).  There
are zero divisors:
\begin{displaymath}
  \rmh_\pm=\textstyle\frac{1}{\sqrt{2}}(1\pm j), \qquad\text{ such that }\quad
  \rmh_+ \rmh_-=0 
  \quad
  \text{ and }
  \quad
  \rmh_\pm^2=\rmh_\pm.
\end{displaymath}
Thus double numbers algebraically isomorphic to two copies of
\(\Space{R}{}\) spanned by \(\rmh_\pm\). Being algebraically dull
double numbers are nevertheless interesting as a homogeneous
space~\cites{Kisil05a,Kisil09c} and they are relevant in
physics~\cites{Khrennikov05a,Ulrych05a,Ulrych08a}.  The combination of
p-mechanical approach with hyperbolic quantum mechanics was already
discussed in~\cite{BrodlieKisil03a}*{\S~6}.

For the hyperbolic character \(\chi_{\rmh \myh}(s)=e^{\rmh \myh
  s}=\cosh \myh s +\rmh\sinh \myh s\)
of \(\Space{R}{}\) one can define
the hyperbolic Fourier-type transform:
\begin{displaymath}
  \hat{k}(q)=\int_{\Space{R}{}} k(x)\,e^{-\rmh q x}dx.
\end{displaymath}
It can be understood in the sense of distributions on the space dual
to the set of analytic functions~\cite{Khrennikov08a}*{\S~3}. Hyperbolic
Fourier transform intertwines the derivative \(\frac{d}{dx}\) and
multiplication by \(\rmh q\)~\cite{Khrennikov08a}*{Prop.~1}.
\begin{example}
  For the Gaussian the hyperbolic Fourier transform is the ordinary
  function (note the sign  difference!):
  \begin{displaymath}
    \int_{\Space{R}{}} e^{-x^2/2} e^{-\rmh q x}dx= \sqrt{2\pi}\, e^{q^2/2}.
  \end{displaymath}
  However the opposite identity:
  \begin{displaymath}
    \int_{\Space{R}{}} e^{x^2/2} e^{-\rmh q x}dx= \sqrt{2\pi}\, e^{-q^2/2}
  \end{displaymath}
  is true only in a suitable distributional sense. To this end we may
  note that \(e^{x^2/2}\) and \(e^{-q^2/2}\) are null solutions to the
  differential operators \(\frac{d}{dx}-x\) and \(\frac{d}{dq}+q\)
  respectively, which are intertwined (up to the factor \(\rmh\)) by
  the hyperbolic Fourier transform. The above differential operators
  \(\frac{d}{dx}-x\) and \(\frac{d}{dq}+q\) are images of the \emph{ladder
  operators} in the Lie algebra of the Heisenberg group. They are
  intertwining by the Fourier transform, since this is an automorphism
  of the Heisenberg group~\cite{Howe80a}.  A careful study of ladder
  operators reveals connections with hypercomplex
  numbers~\cite{Kisil11a,Kisil11c}.
\end{example}
An elegant theory of hyperbolic Fourier transform may be achieved by a
suitable adaptation of~\cite{Howe80a}, which uses representation
theory of the Heisenberg group.

\subsubsection{Hyperbolic Representations of the Heisenberg Group} 
\label{sec:segre-quatern-hyperb}

Consider the space
\(\FSpace[\rmh]{F}{\myh}(\Space{H}{n})\) of \(\Space{O}{}\)-valued
functions on \(\Space{H}{n}\) with the property:
\begin{equation}
  \label{eq:induced-prop-h}
  f(s+s',h,y)=e^{\rmh \myh s'} f(s,x,y), \qquad \text{ for all }
  (s,x,y)\in \Space{H}{n},\ s'\in \Space{R}{} ,
\end{equation}
and the square integrability condition~\eqref{eq:L2-condition}. Then
the hyperbolic representation is obtained by the restriction of the
left shifts to \(\FSpace[\rmh]{F}{\myh}(\Space{H}{n})\).
To obtain an equivalent representation on the phase space we take
\(\Space{O}{}\)-valued functional of the Lie algebra \(\algebra{h}_n\): 
\begin{equation}
  \label{eq:hyp-character}
  \chi^j_{(\myh,q,p)}(s,x,y)=e^{\rmh(\myh s +qx+ py)}
  =\cosh (\myh s +qx+ py) + \rmh\sinh(\myh s +qx+ py).
\end{equation}
The hyperbolic Segal---Bargmann type representation is intertwined
with the left group action by means of the Fourier
transform~\eqref{eq:fourier-transform} with the hyperbolic
functional~\eqref{eq:hyp-character}. Explicitly this representation is:
\begin{equation}
  \label{eq:segal-bargmann-hyp}
  \uir{}{\myhbar}(s,x,y): f (q,p) \mapsto 
  \textstyle e^{-\rmh(\myh s+qx+py)}
  f \left(q-\frac{\myh}{2} y, p+\frac{\myh}{2} x\right).
\end{equation}
For a hyperbolic Schr\"odinger type representation we again use the
scheme described in Rem.~\ref{re:induced-hs}. Similarly to the
elliptic case one obtains the formula,
resembling~\eqref{eq:schroedinger-rep}:
\begin{equation}
    \label{eq:schroedinger-rep-hyp}
    [\uir{\rmh}{\chi}(s',x',y') f](x)=e^{-\rmh\myh (s'+xy'-x'y'/2)}f(x-x').
\end{equation}
Application of the hyperbolic Fourier transform produces a
Schr\"odinger type representation on the configuration space,
cf.~\eqref{eq:schroedinger-rep-conf}: 
\begin{equation}
  \label{eq:schroedinger-rep-conf-hyp}
    [\uir{\rmh}{\chi}(s',x',y') \hat{f}\,](q)=e^{-\rmh\myh (s'+x'y'/2)
    -\rmh x' q}\,\hat{f}(q+\myh y').  
\end{equation}
The extension of this representation to kernels according
to~\eqref{eq:rho-extended-to-L1} generates hyperbolic
pseudodifferential operators introduced
in~\cite{Khrennikov08a}*{(3.4)}.

\subsubsection{Hyperbolic Dynamics}
\label{sec:hyperbolic-dynamics}

Similarly to the elliptic (quantum) case we consider a convolution
of two kernels on \(\Space{H}{n}\) restricted to
\(\FSpace[\rmh]{F}{\myh}(\Space{H}{n})\). The composition law becomes,
cf.~\eqref{eq:composition-ell}:
\begin{equation}
  \label{eq:composition-par}
  (k'*k)\hat{_s}%(-\myh,x,y) 
  =
  \int_{\Space{R}{2n}} e^{ {\rmh \myh}{}(xy'-yx')}\, \hat{k}'_s(\myh ,x',y')\,
 \hat{k}_s(\myh ,x-x',y-y')\,dx'dy'. 
\end{equation}
This is close to the calculus of hyperbolic PDO obtained
in~\cite{Khrennikov08a}*{Thm.~2}.
Respectively for the commutator of two convolutions we get,
cf.~\eqref{eq:repres-commutator}:
\begin{equation}
  \label{eq:commut-par}
  [k',k]\hat{_s}%(-\myh,x,y) 
  = 
  \int_{\Space{R}{2n}}\!\! \sinh(\myh
   (xy'-yx'))\, \hat{k}'_s(\myh ,x',y')\,
 \hat{k}_s(\myh ,x-x',y-y')\,dx'dy'. 
\end{equation}
This the hyperbolic version of the Moyal bracket,
cf.~\cite{Khrennikov08a}*{p.~849}, which generates the corresponding
image of the dynamic equation~\eqref{eq:universal}.
\begin{example}
  \begin{enumerate}
  \item 
    \label{it:hyp-harm-oscil} 
    For a quadratic Hamiltonian, e.g.  harmonic oscillator from
    Example~\ref{ex:p-harmonic}, the hyperbolic equation and
    respective dynamics is identical to quantum considered before.
  \item Since \(\frac{\partial}{\partial s}\) acts on
    \(\FSpace[\rmh]{F}{2}(\Space{H}{n})\) as multiplication by \(\rmh
    \myh\) and \(\rmh^2=1\), the hyperbolic image of the unharmonic
    equation~\eqref{eq:p-unharm-osc-dyn} becomes:
    \begin{displaymath}
      \dot{f}=     \left(m \omega^2 q\frac{\partial}{\partial
          p}+\frac{\lambda}{6}\left(3q^2\frac{\partial}{\partial p}
          +\frac{\myhbar^2}{4}\frac{\partial^3}{\partial p^3}\right)-\frac{1}{m}
        p \frac{\partial}{\partial q} \right) f. 
    \end{displaymath}
    The difference with quantum mechanical
    equation~\eqref{eq:q-unhar-dyn} is in the sign of the cubic
    derivative. 
  \end{enumerate}
\end{example}

\subsubsection{Hyperbolic Probabilities}
\label{sec:hyperb-prob}
\begin{figure}[htbp]
  \centering
  (a)\includegraphics[scale=.75]{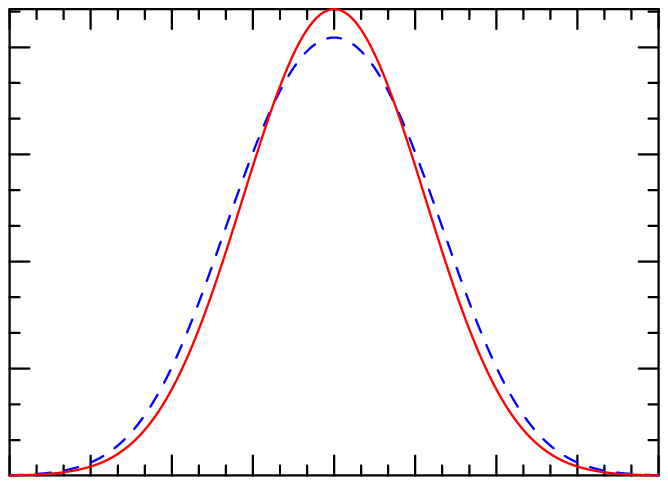}\hfill
  (b)\includegraphics[scale=.75]{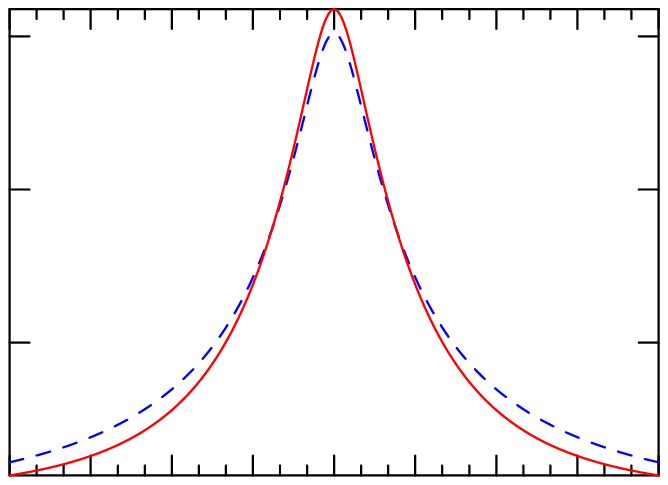}
  \caption{Hyperbolic probabilities: the blue (dashed) graph shows the
    addition of probabilities without interaction, the red (solid)
    graph present the quantum interference. Left picture shows the
    Gaussian state~\eqref{eq:gauss-state}, with the same distribution
    as in quantum mechanics, cf.~Fig.~\ref{fig:quant-prob}(a). The
    right picture shows the rational
    state~\eqref{eq:poly-state}, note the absence of interference
    oscillations in comparison with the quantum
    state on~Fig.~\ref{fig:quant-prob}(b).} 
  \label{fig:hyp-prob}
\end{figure}

To calculate probability distribution generated by a hyperbolic state we are
using the general procedure from
Section~\ref{sec:states-probability}. The main differences with the
quantum case are as follows:
\begin{enumerate}
\item The real number \(A\) in the
  expression~\eqref{eq:addition-functional} for the addition of
  probabilities is bigger than \(1\) in absolute value by. Thus it can
  be associated with the hyperbolic cosine \(\cosh \alpha \),
  cf.~Rem.~\ref{re:sine-cosine}, for certain phase
  \(\alpha\in\Space{R}{}\)~\cite{Khrennikov08a}.
\item The nature of hyperbolic interference on two slits is affected
  by the fact that \(e^{\rmh \myh s}\) is not periodic and the
  hyperbolic exponent \(e^{\rmh t}\) and cosine \(\cosh t\) do not
  oscillate. It is worth to notice that for Gaussian states the
  hyperbolic interference is exactly the same as quantum one,
  cf.~Figs.~\ref{fig:quant-prob}(a) and~\ref{fig:hyp-prob}(a). This is
  similar to coincidence of quantum and hyperbolic dynamics of
  harmonic oscillator.

  The contrast between two types of interference is prominent for
  the rational state~\eqref{eq:poly-state}, which is far from the
  minimal uncertainty, see the different patterns
  on Figs.~\ref{fig:quant-prob}(b) and~\ref{fig:hyp-prob}(b).
\end{enumerate}

\subsection{Parabolic (Classical) representations on the phase space}
\label{sec:class-repr-phase}
After the previous two cases it is natural to link classical mechanics
with dual numbers generated by the parabolic unit \(\rmp^2=0\).
Connection of the parabolic unit \(\rmp\) with the Galilean group of
symmetries of classical mechanics is around for a
while~\cite{Yaglom79}*{App.~C}. 

However the nilpotency of the parabolic unit \(\rmp\) make
it difficult if we will work with dual number valued functions only.
To overcome this issue we consider a commutative real algebra
\(\algebra{C}\) spanned by \(1\), \(\rmi\), \(\rmp\) and \(\rmi\rmp\)
with identities \(\rmi^2=-1\) and \(\rmp^2=0\). A seminorm on
\(\algebra{C}\) is defined as follows:
\begin{displaymath}
  \modulus{a+b\rmi+c\rmp+d\rmi\rmp}^2=a^2+b^2.
\end{displaymath}

\subsubsection{Classical Non-Commutative Representations}
\label{sec:class-non-comm}
We wish to build a representation of the Heisenberg group which will
be a classical analog of the Segal--Barg\-mann
representation~\eqref{eq:stone-inf}.  To this end we introduce the
space
\(\FSpace[\rmp]{F}{\myh}(\Space{H}{n})\) of \(\algebra{C}\)-valued
functions on \(\Space{H}{n}\) with the property:
\begin{equation}
  \label{eq:induced-prop-p}
  f(s+s',h,y)=e^{\rmp \myh s'} f(s,x,y), \qquad \text{ for all }
  (s,x,y)\in \Space{H}{n},\ s'\in \Space{R}{} ,
\end{equation}
and the square integrability condition~\eqref{eq:L2-condition}. It is
invariant under the left shifts and we restrict the left group action to
\(\FSpace[\rmp]{F}{\myh}(\Space{H}{n})\). 

There is an unimodular
\(\algebra{C}\)-valued function on the Heisenberg group parametrised
by a point \((\myh, q, p)\in\Space{R}{2n+1}\):
\begin{displaymath}
  E_{(\myh,q,p)}(s,x,y)= e^{2\pi(\rmp s\myhbar+\rmi xq + \rmi yp)}=e^{2\pi\rmi (xq + yp)}(1+\rmp s\myh).
\end{displaymath}
This function, if used instead of the ordinary exponent, produces a modification
\(\oper{F}_c\) of the Fourier transform~\eqref{eq:fourier-transform}.
The transform intertwines the left regular representation with the following
action on \(\algebra{C}\)-valued functions on the phase space:
\begin{equation}
  \label{eq:dual-repres}
  \uir{\rmp}{\myh}(s,x,y): f(q,p) \mapsto e^{-2\pi\rmi(xq+yp)}(f(q,p)
  +\rmp\myh(s f(q,p) +\frac{y}{2\pi\rmi}f'_q(q,p)-\frac{x}{2\pi\rmi}f'_p(q,p))).
\end{equation}
\begin{rem}
  \label{re:classic-rep}
  Comparing the traditional
  infinite-dimensional~\eqref{eq:stone-inf} and
  one-dimen\-sional~\eqref{eq:commut-repres} representations of
  \(\Space{H}{n}\) we can note that the properties of the
  representation~\eqref{eq:dual-repres} are a non-trivial mixture of
  the former:  
  \begin{enumerate}
  \item \label{it:class-non-commut} 
    The action~\eqref{eq:dual-repres}
    is non-commutative, similarly to the quantum
    representation~\eqref{eq:stone-inf} and unlike the classical
    one~\eqref{eq:commut-repres}. This non-commutativity will produce
    the Hamilton equations below in a way very similar to Heisenberg
    equation, see Rem.~\ref{re:hamilton-from-nc}.
  \item \label{it:class-locality} The
    representation~\eqref{eq:dual-repres} does not change the support
    of a function \(f\) on the phase space, similarly to the
    classical representation~\eqref{eq:commut-repres} and unlike the
    quantum one~\eqref{eq:stone-inf}. Such a localised action will be
    responsible later for an absence of an interference in classical
    probabilities.
  \item The parabolic representation~\eqref{eq:dual-repres} can not be
    derived from either the elliptic~\eqref{eq:stone-inf} or
    hyperbolic~\eqref{eq:segal-bargmann-hyp} by the plain substitution
    \(\myh=0\).
  \end{enumerate}
\end{rem}
We may also write a classical Schr\"odinger type representation.
According to Rem.~\ref{re:induced-hs} we get a representation formally
very similar to the elliptic~\eqref{eq:schroedinger-rep} and
hyperbolic versions~\eqref{eq:schroedinger-rep-hyp}:
\begin{eqnarray}
    \label{eq:schroedinger-rep-par}
    [\uir{\rmp}{\chi}(s',x',y') f](x)&=&e^{-\rmp\myh
      (s'+xy'-x'y'/2)}f(x-x')\\
    &=&(1-\rmp\myh (s'+xy'-\textstyle\frac{1}{2}x'y')) f(x-x').\nonumber 
\end{eqnarray}
However due to nilpotency of \(\rmp\) the (complex) Fourier transform
\(x\mapsto q\) produces a different formula for parabolic
Schr\"odinger type representation in the configuration space,
cf.~\eqref{eq:schroedinger-rep-conf}
and~\eqref{eq:schroedinger-rep-conf-hyp}:
\begin{equation}
  \label{eq:schroedinger-rep-conf-par}
    [\uir{\rmp}{\chi}(s',x',y') \hat{f}](q)= e^{2\pi\rmi x' q}\left(
    \left(1-\rmp\myh (s'-{\textstyle\frac{1}{2}}x'y')\right)    \hat{f}(q)
    +\frac{\rmp\myh y'}{2\pi\rmi}\hat{f}'(q)\right).
\end{equation}
This representation shares all properties mentioned in
Rem.~\ref{re:classic-rep} as well.

\subsubsection{Hamilton Equation}
\label{sec:hamilton-equation}

% The convolution of two functions \(e^{\rmp \myh s}k_i(x,y)\),
% \(i=1\), \(2\) from \(\FSpace{F}{\myh}\) is 
% \begin{eqnarray}
%   \lefteqn{k_1*k_2(s,x,y) 
%   = 
%   c_\myh^{n+1} \int_{\Space{H}{n}}e^{\rmp \myh s'} k_1(x',y') }
%  \label{eq:almost-star-product}\\%*
%  &&{}\times e^{\rmp \myh (s-s'+\frac{1}{2}%*
%  (xy'-yx'))} k_2(x-x',y-y')\,ds'dx'dy' \nonumber \\%*
%   &=& 
%   e^{\rmp \myh s}c_\myh^{n+1} \int_{\Space{H}{n}} e^{ \frac{\rmp \myh}{2}%*
%  (xy'-yx')} \, k_1(x',y') \nonumber 
%  \label{eq:almost-star-product}\\%*
%  &&\quad \qquad {} \times k_2(x-x',y-y')\,ds'dx'dy' \nonumber 
% \end{eqnarray}

The identity \(e^{ \rmp t}-e^{ -\rmp t}= 2\rmp t\) can be interpreted
as a parabolic version of the sine function, while the parabolic
cosine is identically equal to
one~\cites{HerranzOrtegaSantander99a,Kisil07a}.  From this we obtain
the parabolic version of the commutator~\eqref{eq:repres-commutator}:
\begin{eqnarray*}
  [k',k]\hat{_s}(\rmp \myh, x,y) 
  &=& 
  %e^{\rmp \myh s}c_\myh^{n+1}  
  \rmp \myh\int_{\Space{R}{2n}}
 (xy'-yx') \\
 &&{}\times\, \hat{k}'_s(\rmp \myh,x',y')  \,
 \hat{k}_s(\rmp \myh,x-x',y-y')\,dx'dy', \nonumber 
\end{eqnarray*}
for the partial parabolic Fourier-type transform \(\hat{k}_s\) of the
kernels.  Thus the parabolic representation of the dynamical
equation~\eqref{eq:universal} becomes:
\begin{equation}
  \label{eq:dynamics-par}
  \rmp\myh \frac{d\hat{f}_s}{dt}(\rmp\myh,x,y;t)=
  %e^{\rmp \myh s}c_\myh^{n+1} 
  \rmp \myh \int_{\Space{R}{2n}}
 (xy'-yx')\, %\\
 %&&{}\times\, 
\hat{H}_s(\rmp \myh,x',y')  \,
 \hat{f}_s(\rmp \myh,x-x',y-y';t)\,dx'dy', %\nonumber 
\end{equation}
Although there is no possibility to divide by \(\rmp\) (since it is a
zero divisor) we can obviously eliminate \(\rmp \myh \) from the both
sides if the rest of the expressions are real.  Moreover this can be
done ``in advance'' through a kind of the antiderivative operator
considered in~\cite{Kisil02e}*{(4.1)}. This will prevent ``imaginary
parts'' of the remaining expressions (which contain the factor
\(\rmp\)) from vanishing.
\begin{rem}
  It is noteworthy that the Planck constants completely disappeared
  from the dynamical equation. Thus the only prediction about it
  following from our construction is \(\myh\neq 0\), which was
  confirmed by experiments, of course. 
\end{rem}
Using the duality between the Lie algebra of \(\Space{H}{n}\) and the
phase space we can find an adjoint equation for observables on the
phase space. To this end we apply the usual Fourier transform
\((x,y)\mapsto(q,p)\). It turn to be the Hamilton
equation~\cite{Kisil02e}*{(4.7)}.  However the transition to phase
space is more a custom rather than a necessity and in many cases we
can efficiently work on the Heisenberg group itself.

\begin{rem}
  \label{re:hamilton-from-nc}
  It is noteworthy, that the non-commutative
  representation~\eqref{eq:dual-repres} allows to obtain the Hamilton
  equation directly from the commutator
  \([\uir{\rmp}{\myh}(k_1),\uir{\rmp}{\myh}(k_2)]\). Indeed its
  straightforward  evaluation will produce exactly the above expression. On
  the contrast such a commutator for the commutative
  representation~\eqref{eq:commut-repres} is zero and to obtain the
  Hamilton equation we have to work with an additional tools, e.g. an
  anti-derivative~\cite{Kisil02e}*{(4.1)}. 
\end{rem}

\begin{example}
  \begin{enumerate}
  \item For the harmonic oscillator in Example~\ref{ex:p-harmonic} the
    equation~\eqref{eq:dynamics-par} again reduces to the
    form~\eqref{eq:p-harm-osc-dyn} with the solution given
    by~\eqref{eq:p-harm-sol}. The adjoint equation of the harmonic
    oscillator on the phase space is not different from the quantum
    written in Example~\ref{it:q-harmonic}. This is true for any
    Hamiltonian of at most quadratic order.
  \item 
    For non-quadratic Hamiltonians classical and quantum dynamics
    are different, of course. For example, 
    the cubic term of \(\partial_s\) in the
    equation~\eqref{eq:p-unharm-osc-dyn} will generate the factor
    \(\rmp^3=0\) and thus vanish. Thus the
    equation~\eqref{eq:dynamics-par} of the unharmonic oscillator on
    \(\Space{H}{n}\) becomes:
    \begin{displaymath}
      %\label{eq:p-unharm-osc-dyn-par}
      \dot{f}=     \left(m \omega^2 y\frac{\partial}{\partial x}
        +\frac{\lambda y}{2}\frac{\partial^2}{\partial x^2} 
          -\frac{1}{m} x
        \frac{\partial}{\partial y} \right) f. 
   \end{displaymath}
   The adjoint equation on the phase space is:
    \begin{displaymath}
      \dot{f}=     \left(\left(m \omega^2
          q+\frac{\lambda}{2}q^2\right)
        \frac{\partial}{\partial p} -\frac{1}{m}  p \frac{\partial}{\partial q} \right) f. 
    \end{displaymath}
    The last equation is the classical
    Hamilton equation generated by the
    cubic potential~\eqref{eq:unharmonic-hamiltonian}. Qualitative
    analysis of its dynamics can be found in many textbooks
    \citelist{\cite{Arnold91}*{\S~4.C, Pic.~12} \cite{PercivalRichards82}*{\S~4.4}}. 
  \end{enumerate}
\end{example}

\begin{rem}
  We have obtained the Poisson bracket from the commutator of
  convolutions on \(\Space{H}{n}\) without any quasiclassical limit
  \(\myh\rightarrow 0\). This has a common source with the deduction
  of main calculus theorems in~\cite{CatoniCannataNichelatti04} based
  on dual numbers. As explained in~\cite{Kisil05a}*{Rem.~6.9} this is
  due to the similarity between the parabolic unit \(\rmp\) and the
  infinitesimal number used in non-standard analysis~\cite{Devis77}.
  In other words, we never need to take care about terms of order
  \(O(\myh^2)\) because they will be wiped out by \(\rmp^2=0\).
\end{rem}
An alternative derivation of classical dynamics from the Heisenberg
group is given in the recent paper~\cite{Low09a}.

\subsubsection{Classical probabilities}
\label{sec:class-prob}
It is worth to notice that dual numbers are not only helpful in
reproducing classical Hamiltonian dynamics, they also provide the
classic rule for addition of probabilities.
We use the same formula~\eqref{eq:kernel-state} to calculate kernels of
the states. The important difference now that the
representation~\eqref{eq:dual-repres} does not change the support of
functions. Thus if we calculate the correlation term
\(\scalar{v_1}{\uir{}{}(g)v_2}\) in~\eqref{eq:kernel-add}, then it
will be zero for every two vectors \(v_1\) and \(v_2\) which have 
disjoint supports in the phase space. Thus no interference similar to
quantum or hyperbolic cases (Subsection~\ref{sec:quantum-probabilities})
is possible.

\section{Discussion}
\label{sec:discussion}

In this paper we derive mathematical models for various physical setup
from hypercomplex representations of the Heisenberg group. There are
roots for such hypercomplex characters in the structure of ladder
operators associated to three non-isomorphic quadratic
Hamiltonians~\cites{Kisil11a,Kisil11c}. Such hypercomplex
representations may be also useful for many other groups as well, see
the example of the \(\SL\) group in~\cite{Kisil09c}.  Moreover
non-trivial parabolic characters described
in~\cites{Kisil07a,Kisil09c} are awaiting a further exploration.

There is a connection of our work with the technique of contractions
and analytic continuations of groups~\cites{Gromov90b,Gromov90a},
these papers also highlight the role of hypercomplex numbers of three
types. However in our research we do not modify the group (the
Heisenberg group more specifically) itself, we rather consider its
representations in different functional spaces created by three types
of hypercomplex characters. All three cases have a lot of algebraic
similarity and can be written in a unified manner with the help of
parameter, which takes three values, say \(u=\rmi\), \(\rmp\),
\(\rmh\), with \(\rmi^2=-1\), \(\rmp^2=0\), \(\rmh^2=1\). For example,
representations~\eqref{eq:schroedinger-rep},
\eqref{eq:schroedinger-rep-hyp} and~\eqref{eq:schroedinger-rep-par}
can be unified in:
\begin{equation}
  \label{eq:schroedinger-unified}
  [\uir{u}{\myh}(s',x',y')f](x)=e^{-u\myh(s'+xy'-x'y'/2)}f(x-x').
\end{equation}
It is noteworthy that this algebraic similarity exists along with the
significant topological and analytic differences between elliptic,
parabolic and hyperbolic cases. An illustration is the distinction of
the elliptic~\eqref{eq:schroedinger-rep-conf} and
parabolic~\eqref{eq:schroedinger-rep-conf-par} representations in the
configuration space, despite of the fact that both representations are
derived from the unified form~\eqref{eq:schroedinger-unified}.

The parabolic representations~\eqref{eq:schroedinger-rep-par}
and~\eqref{eq:schroedinger-rep-conf-par} of the Heisenberg group act
in the first order jet spaces. Such spaces have a well established
connections with Lagrangian and Hamiltonian formulations of quantum
field
theory~\cites{GiachettaMangiarottiSardanashvily97a,Kanatchikov01b,Kisil04a},
study of aggregate quantum-classical systems~\cites{Kisil05c,Kisil09a}
and spectral theory of operators~\cite{Kisil02a}. Nevertheless the
localised non-commutative representation of \(\Space{H}{n}\) built in
this paper seems to be new and deserve detailed investigation.

We already seen that it may be useful to consider several hypercomplex
units in the same time.  In the case of classical mechanics we
combined \(\rmi\) and \(\rmp\). The algebra generated by \(\rmi\) and
\(\rmh\) is known as (commutative) Segre quaternions. Such commutative
algebras with hypercomplex units and their physical applications
attracted attention of many researchers
recently~\citelist{\cite{BocCatoniCannataNichZamp07} \cite{Plaksa09a}
  \cite{Ulrych05a} \cite{Ulrych08a}}. 

We may even need to study an algebra which contains all three
hypercomplex units simultaneously. The most straightforward way is to
take eight dimensional commutative algebra with the basis \(1\),
\(\rmi\), \(\rmp\), \(\rmh\), \(\rmi\rmp\), \(\rmi\rmh\),
\(\rmp\rmh\), \(\rmi\rmp\rmh\). A reduction of dimensionality from
\(8\) to \(6\) can be achieved if we replace products \(\rmp\rmh\) and
\(\rmi\rmp\rmh\) through the further identities \(\rmp\rmh=\rmp\) and
\(\rmi\rmp\rmh=\rmi\rmp\). This do not affect associativity of the
product.

\section*{Acknowledgements}
\label{sec:acknowledgments}
I am grateful to A.Yu.~Khrennikov and S.~Ulrych for useful discussion
on relation between double numbers and physics. S.~Plaksa advised me
on various aspect of commutative hypercomplex algebras. U.~G\"uenther
draw my attention to the connection between \(\mathcal{PT}\)-symmetric
Hamiltonians and Krein spaces. Prof. N.A.~Gromov made several useful
suggestions of methodological nature. Constructive comments of
anonymous referees provided further ground for paper's improvement.

%\label{se:appendix}

\small
\bibliography{abbrevmr,akisil,analyse,algebra,arare,aclifford,aphysics,ageometry}
\end{document}